\documentclass[twocolumn,secnumarabic,amssymb, nobibnotes, aps, pre]{revtex4-2}

\usepackage{framed}
\usepackage{amsfonts}
\usepackage{amsmath}
\usepackage{graphicx}
\usepackage{color}

\usepackage[hidelinks]{hyperref}
\usepackage{xcolor}
\hypersetup{
	colorlinks,
	linkcolor={blue!80},
	citecolor={blue},
	urlcolor={blue!80!black}
}

\begin{document}
	\title{Pure quartic three-dimensional spatiotemporal Kerr solitons }
	
	\author{Pedro Parra-Rivas$^*$, Yifan Sun, Fabio Mangini, Mario Ferraro, Mario Zitelli, Stefan Wabnitz}
	\email{pedro.parra-rivas@uniroma1.it}
	\affiliation{
		Dipartimento  di  Ingegneria  dell’Informazione, Elettronica  e  Telecomunicazioni,
		Sapienza  Universit{\'a}  di  Roma, via  Eudossiana  18, 00184  Rome, Italy\\
	}


\begin{abstract}
We analyze the formation of three-dimensional spatiotemporal solitons in waveguides with a parabolic refractive index profile and pure quartic chromatic dispersion. We show, by applying both variational approaches and full three-dimensional numerical simulations, that fourth-order dispersion has a positive impact on soliton stabilization against spatiotemporal wave collapse. Specifically, pure quartic spatiotemporal solitons remain stable within a significantly larger energy range with respect to their second-order dispersion counterparts. 
\end{abstract}

\maketitle
The formation of high-dimensional solitons is a very intense field of research in different domains of science, ranging from nonlinear optics to Bose-Einstein condensates (BECs) \cite{kivshar_optical_2003,kartashov_frontiers_2019,10.1063/9780735425118}. In nonlinear optics, the formation of three-dimensional spatiotemporal solitons (STS), also known as {\it light bullets} \cite{silberberg_collapse_1990}, has been predicted in 
Kerr nonlinear lossless materials. Their mechanism is a counterbalance between the action of the intensity-dependent refractive index on the one hand, and the combined effect of dispersion and diffraction on the other hand \cite{kivshar_optical_2003}. One of the main properties of these nonlinear waves is that, once they form, they can propagate indefinitely, without any shape modification. However, in experiments, the STSs have only been generated as transient objects, owing to instabilities associated with the presence of high-order effects \cite{PhysRevLett.105.263901,renninger_optical_2013,panagiotopoulos_super_2015}.

In Kerr media, the most common instability is {\it spatiotemporal wave collapse}, whereby
the strong contraction of a nonlinear wave leads to a catastrophic blow-up of its amplitude after a finite propagation distance 
\cite{silberberg_collapse_1990,berge_wave_1998,bang_collapse_2002}.  Many different mechanisms have been proposed for STS stabilization, including saturable absorption, nonlocal and quadratic nonlinearities, and photonic lattices, to cite a few \cite{kartashov_frontiers_2019,10.1063/9780735425118}. Another stabilization mechanism relies on a parabolic modification of the transverse refractive index profile of the material, as it occurs in graded index (GRIN) multimode waveguides or fibers \cite{horak_multimode_2012}.

The parabolic index profile acts as a trapping potential, which is able to arrest the wave collapse, as it was predicted by Yu {\it et al.} \cite{yu_spatio-temporal_1995} and Raghavan {\it et al.} \cite{raghavan_spatiotemporal_2000} by means of variational approaches. For low pulse energy regimes, these results are well confirmed by full 3D numerical solutions of the nonlinear wave equation.
However, for sufficiently high pulse energies,
even below the theoretical predicted stability threshold, this mechanism fails to arrest the collapse \cite{parrarivas2023dynamics}. Thus, one may wonder if there are other alternatives for enlarging the energy-dependent stability range of light bullets.

The use of high-order dispersive effects has proved key for stabilizing temporal solitons in nonlinear cavities: specifically, consider the case of third- \cite{milianOE,parra-rivas_third-order_2014,parra-rivas_coexistence_2017} or fourth-order dispersion \cite{tlidi_high-order_2010}. Moreover, the effect of pure quartic dispersion in soliton formation has been studied in the context of micro-comb generation \cite{tahOL19,Parra-Rivas:22}, mode-locked lasers \cite{redNP}, and single-pass (conservative) systems, where pure quartic solitons have been theoretically studied \cite{karlsson_soliton-like_1994,akhmediev_radiationless_1994,PhysRevA.87.025801,Tam:19} and experimentally demonstrated \cite{karlsFOD93,redNC,de_sterke_pure-quartic_2021}. Pure quartic temporal solitons possess flatter spectra and favorable energy scaling with pulse duration, which makes them particularly attractive from the point of view of applications.

In this letter, we demonstrate the formation of pure quartic STSs in GRIN waveguides, and show that pure quartic dispersion alone is able to significantly suppresses wave collapse, thus greatly favoring STS stability. To show this, we follow a two-fold approach, based on both the Ritz optimization method (i.e., the variational approach) \cite{perez-garcia_dynamics_1997,malomed_variational_2002}, and direct full 3D numerical simulations.

\begin{figure}[!t]
	\centering
	\includegraphics[scale=1]{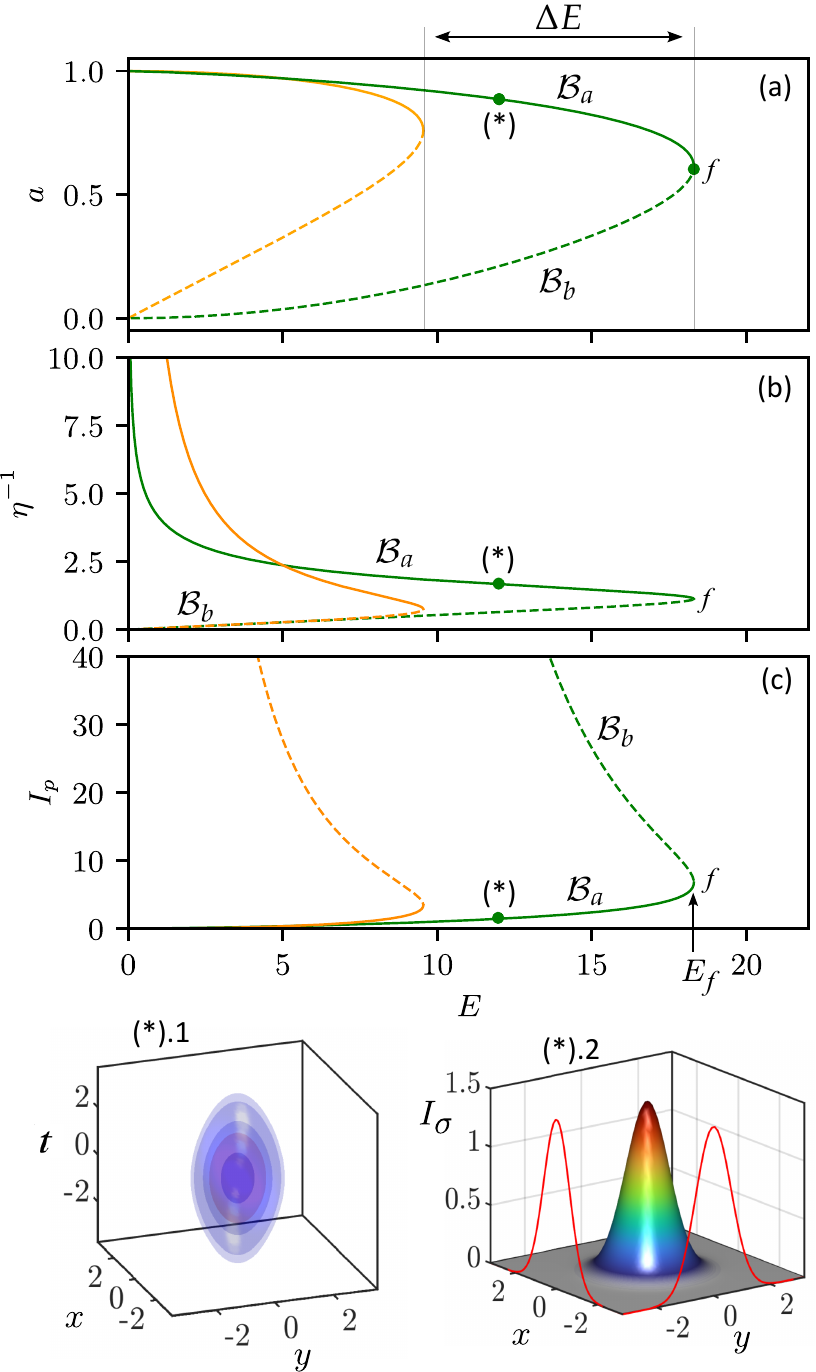}
	\caption{Bifurcation diagrams for STS states vs. $E$ (see green lines). In (a) the spatial width $a$ of the STS vs. $E$ is depicted. Panel (b) shows the STS temporal width $\eta^{-1}$, and panel (c) shows the variation of the peak intensity $I_p$ with $E$. The $\mathcal{B}_a$ branch of solutions is plotted by using a solid line, while $\mathcal{B}_b$ uses a dashed one. Orange lines correspond to the pure quadratic scenario \cite{parrarivas2023dynamics}.  Label (*) corresponds to the STS depicted below in panels (*).1 and (*).2. In (*).1 we plot five isosurfaces at different peak intensities, namely $I=0.08,0.12.0.3,0.5,1.0$. Panel (*).2 illustrates the $t=0$ cross-section intensity $I_\sigma\equiv I(x,y,t=0)$ for the STSs shown above.
    }
	\label{fig1}
\end{figure}
In the paraxial and slowly varying envelope approximations, the amplitude of the electric field $E$ propagating in waveguides with a GRIN profile at the carrier frequency $\omega_0$ can be described by the Gross-Pitaevskii equation with a 2D parabolic potential
\begin{equation}\label{Eq_first}
	\partial_Z E=\frac{i}{2\beta_0}\nabla_\perp^2E+i\frac{\beta_4}{4!}\partial_T^4 E+i\frac{n_1}{k_0}(X^2+Y^2)E+ik_0n_2|E|^2E
\end{equation} 
where $k_0=\omega_0/c$, $\beta_0=n_0(\omega_0)k_0$, $\beta_4=d^4\beta/d\omega^4|_{\omega_0}$, and $\beta(\omega)=n_0(\omega)\omega/c$,  with $n_0(\omega)$ being the homogeneous contribution of the refractive index. $\nabla_\perp^2\equiv\partial_X^2+\partial_Y^2$ represents diffraction, the $\partial_t^4$ term results from fourth-order chromatic dispersion, and $n_2$ is the refractive index nonlinear coefficient responsible for the self-focusing or self-defocusing Kerr nonlinearity \cite{kivshar_optical_2003,horak_multimode_2012}. 
By taking the scaling transformations $E=e_cu$, $T=t_ct$, $(X,Y)=w_c(x,y)$ and $Z=z_cz$ with $e_c^4\equiv2|n_1|/(|n_2|^2\beta_0 k_0^3)$, $t_c^4\equiv (|\beta_4|/4!)\sqrt{k_0\beta_0/(2|n_1|)}$, $w_c^4\equiv k_0/(2|n_1|\beta_0)$, and $z_c^2\equiv k_0\beta_0/(2|n_1|)$, 
Eq.~(\ref{Eq_first}) becomes  
\begin{equation}\label{GPE}
	\partial_z u=\frac{i}{2}\nabla_\perp^2u+id_4\partial^4_t u+i\frac{\rho}{2}(x^2+y^2)u+i\nu|u|^2u,
\end{equation}
with  $d_4={\rm sign}(\beta_4)=\pm 1$, $\nu={\rm sign}(n_2)=\pm 1$ for self-focusing/self-defocusing nonlinearity, and $\rho={\rm sign}(n_2)=\pm 1$ for anti-guiding/guiding materials, respectively.

The Lagrangian density associated with this equation reads 
\begin{equation}
	\begin{aligned} 
		\mathcal{L}=&-\frac{1}{2}\left(|u_x|^2+|u_y|^2\right)-d_4|u_{tt}|^2+\frac{\rho}{2} (x^2+y^2)|u|^2\\
		&+\frac{\nu}{2} |u|^4 +\frac{i}{2} \left(u^*u_z-u u_z^* \right),
	\end{aligned}
	\label{Lagran1}
\end{equation}
and by defining the generalized field momenta $\mathcal{P}\equiv \partial_{u^*_z}\mathcal{L}=-iu/2$ and $\mathcal{P}^*\equiv \partial_{u_z}\mathcal{L}=iu^*/2$, we can obtain the Hamiltonian density $\mathcal{H}$ through the Legendre transform 
$\mathcal{H}=\mathcal{P} u^*_z+\mathcal{P}^* u_z-\mathcal{L}$ 
 \cite{abraham_foundations_2008}
.   
Here, we focus on shape-preserving and vorticity-free solitons. Therefore, we write  $u(x,y,z,t)=v(x,y,t)e^{i \kappa z}$, where $\kappa$ is the propagation constant (or chemical potential in the context of BECs) \cite{kivshar_optical_2003}, and $v(x,y,t)$ is a real-valued function, describing the steady-state field.
With this transformation, the Lagrangian density becomes 
\begin{equation}
	\begin{aligned} 
		\mathcal{L}_v=-\frac{1}{2}\left(v_x^2+v_y^2\right)-d_4 v_{tt}^2+\frac{\rho}{2} (x^2+y^2)v^2
		+\frac{\nu}{2} v^4-\kappa v^2,
	\end{aligned}
\end{equation}
while the Hamiltonian density reads
\begin{multline}\label{Hamil_density}
	\mathcal{H}=\frac{1}{2}\left( v_x^2+v_y^2\right)+d_4 v_{tt}^2-\frac{\nu}{2}v^4 -\frac{\rho}{2} \left(x^2+y^2\right)v^2. 
\end{multline}
The $z$-independent Euler-Lagrange equations 
\begin{equation}
	\frac{d^2}{dt^2}\left(\frac{\partial\mathcal{L}_v}{\partial v_{tt}}\right)+\frac{d}{dx}\left(\frac{\partial\mathcal{L}_v}{\partial v_x}\right)+\frac{d}{dy}\left(\frac{\partial\mathcal{L}_v}{\partial v_y}\right)-\frac{\partial \mathcal{L}_v}{\partial v}=0.
\end{equation}
lead to the steady-state partial differential equation  
\begin{equation}\label{real_GP}
	\frac{1}{2}\nabla_\perp^2v+d_4 \partial_{t}^4v+\frac{\rho}{2} (x^2+y^2)v+\nu v^3-	\kappa v=0.
\end{equation}
In what follows, by applying the Ritz optimization method \cite{perez-garcia_dynamics_1997,malomed_variational_2002}, we will compute an approximate analytical steady STS solution of Eq.~(\ref{real_GP}). This method relies on the proper selection of a trial function, or solution ansatz. Here, by following previous works \cite{yu_spatio-temporal_1995,raghavan_spatiotemporal_2000,parrarivas2023dynamics}, we consider the parameter-dependent ansatz 
\begin{equation}\label{ansatz1}
	v(x,y,t;\eta,a,E)=\sqrt{\frac{\eta E}{2\pi a^2}}{\rm sech}(\eta t) {\rm Exp}\left(-\frac{x^2+y^2}{2a^2}\right), 
\end{equation}
where $a$ is the width of the spatial Gaussian profile, $\eta^{-1}$ is the temporal width, and $E$ is the STS energy.
\begin{figure}[!t]
	\centering
	\includegraphics[scale=1]{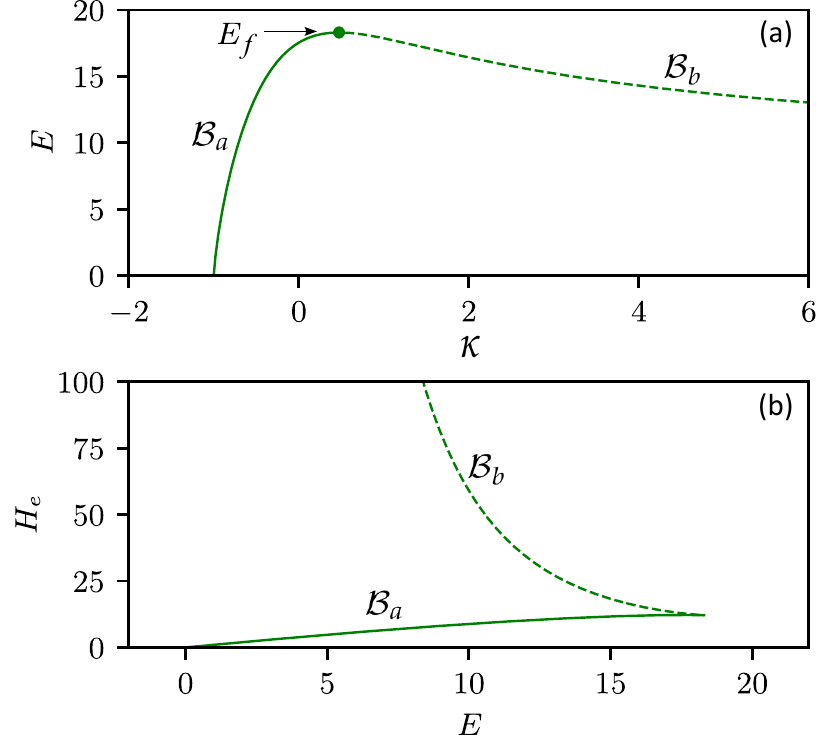}
	\caption{ (a) Dependence of energy $E$ with $\kappa$ for $\rho=-1$ and $\nu=d_4=1$. Panel (b) shows the dependence of $H$ with $E$. In both cases, stable (unstable) branches are plotted by using solid (dashed) lines.
	}
	\label{fig2}
\end{figure}
With this ansatz, the Lagrangian of the system  
\begin{equation}
	L\equiv\int_{{\rm I\!R}^3} \mathcal{L}_v\left(v,v_{tt},v_x,v_y\right)dxdydt
\end{equation}
reduces to
\begin{equation}\label{Lagran}
		L=\frac{E}{30}\left(-14d_4\eta^4-30\kappa+\frac{5}{a^2}\left(\frac{E\eta\nu}{2\pi}-3\right)+15\rho a^2\right),
\end{equation}
\begin{figure*}[!t]
	\centering
	\includegraphics[scale=1]{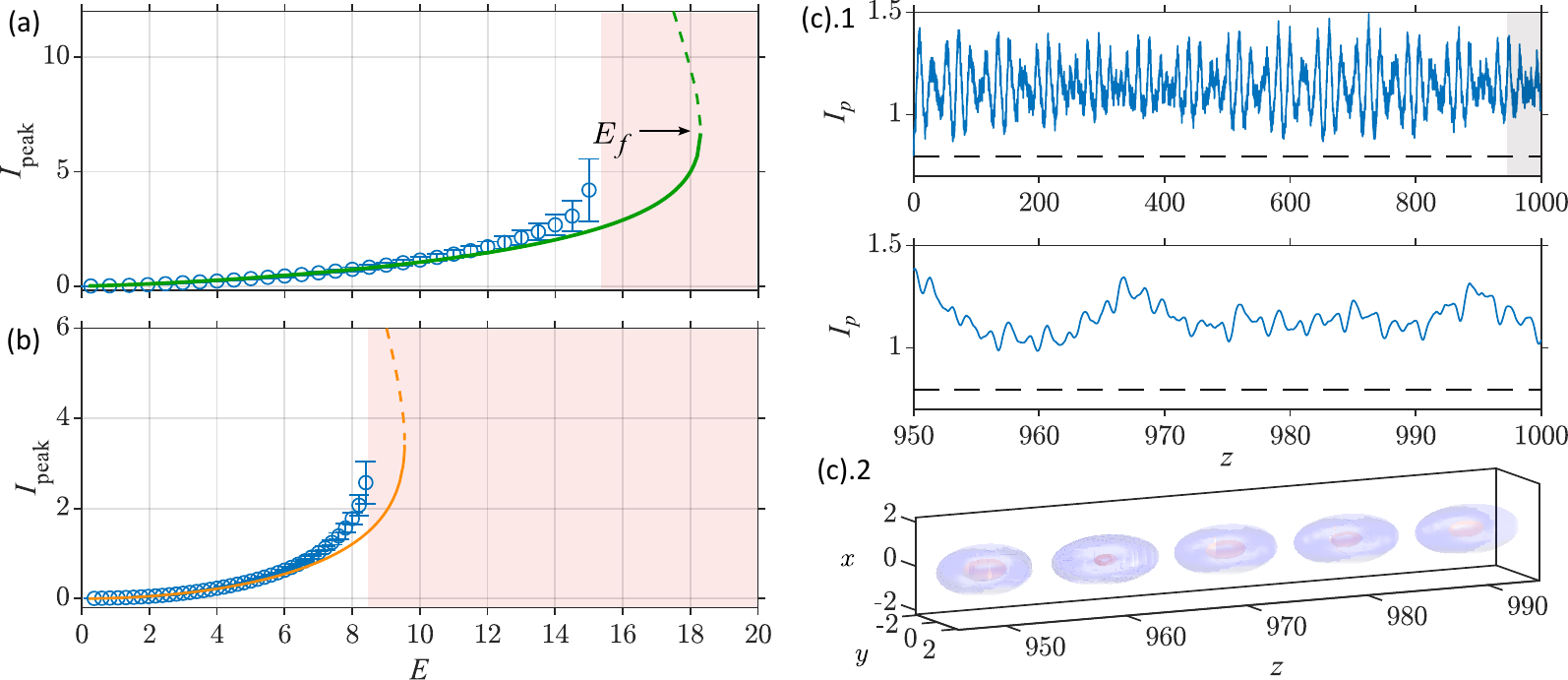}
	\caption{Panels (a) and (b) show the evolution of peak intensity of stable STS with energy $E$ for the pure quartic and quadratic dispersion scenarios, respectively. In (a), the green line shows the analytical values, while the blue circles and the error bars represent the average intensity values, and the standard deviation for stable states. Panel (c).1 shows the variation of the peak STS intensity vs. $z$, and the close-up view below shows the interval $z\in[950,1000]$ (see gray shadowed box). Panel (c).2 shows the evolution of the STS along the latter interval, by plotting two isosurfaces at $I_1=0.5$ (red), and $I_2=0.1$ (blue). }
	\label{fig3}    
\end{figure*} 
\noindent which possesses all relevant information for approximate solutions of the form (\ref{ansatz1}). For Eq.~(\ref{Lagran}), the reduced Euler-Lagrange equations for the parameters $a$, $\eta$ and $E$ read as
\begin{equation}
	\frac{\partial L}{\partial \eta}=0, \qquad \frac{\partial L}{\partial a}=0, \qquad \frac{\partial L}{\partial E}=0,
\end{equation}
which lead, respectively, to the following equations
\begin{subequations}
	\begin{equation}\label{eq1}
		\frac{5 \nu  E }{\pi  a^2}-112 d_4  \eta ^3=0
	\end{equation}
	\begin{equation}\label{eq2}
		\eta=\frac{6 \pi  \left(a^4 \rho +1\right)}{\nu  E }
	\end{equation}
	\begin{equation}\label{eq3}
	\kappa=-\frac{7}{15}d_4\eta^4-\frac{1}{2a^2}\left(1-\frac{E\eta\nu}{3\pi}\right)+\frac{\rho a^2}{2}
	\end{equation}
\end{subequations}

\noindent By combining Eq.~(\ref{eq1}) and Eq.~(\ref{eq2}), we obtain 
\begin{equation}\label{Ener}
	E^4=C_1d_4\pi^4 a^2(a^4\rho+1)^3,
\end{equation}
with $C_1=1008\cdot 24/5$,
which relates $E$ and $a$. By inserting Eq.~(\ref{Ener}) into Eq.~(\ref{eq2}), we find that the temporal width $\eta^{-1}$ is also completely parameterized in terms of the spatial width $a$. 

In what follows, we will focus on the regime that is characterized by setting $d_4=1$, $\rho=-1$, and $\nu=1$. In this case, the dependence of the STS spatial width $a$ on $E$ is depicted in Fig.~\ref{fig1}(a). This plot shows that there exist two STS solutions branches $\mathcal{B}_a$ (solid green) and $\mathcal{B}_b$ (dashed green), which coexist within the same energy range, extending from $E=0$ up to the fold ($f$) located at $(E,a)=(E_f,a_f)$. The position of this fold can be computed from the condition $dE/da=0$, which yields  
\begin{equation}
	a_{f}=(-7\rho)^{-1/4}\qquad E^4_f=C_2\frac{\pi^4 d_4}{\sqrt{-7\rho}},
\end{equation} 
with $C_2=C_1(6/7)^3$,
and marks an upper energy limit, or threshold, for the STS existence. The modification of the temporal width and the STS peak intensity $I_{p}\equiv|A|^2=E\eta/(2\pi a^2)$ with energy $E$ are illustrated in Figs.~\ref{fig1}(b) and \ref{fig1}(c). A specific example of STS solution on the branch $\mathcal{B}_a$ is shown in Fig.~\ref{fig1}(*) for $E=12$. In Figs.~\ref{fig1}(a)-(c)
we also plot, in orange, the STS solution branches for the pure quadratic dispersion regime \cite{parrarivas2023dynamics}. The comparison between these curves shows that
the STS existence region for pure quartic dispersion is $\Delta E\approx 8.753$ larger than in the quadratic case [see Fig.~\ref{fig1}(a)]. 

In order to determine the stability of the STS states, we apply two different approaches. The first, known as the Vakhitov-Kolokov (VK) criterion \cite{vakhitov_stationary_1973}, is based on the dependence of the propagation constant $\kappa$ on $E$ [see Eq.~(\ref{eq3})], which is depicted in Fig.~\ref{fig2}(a). According to the VK principle, STS solutions are stable if $E$ increases with $\kappa$ (i.e., if $dE/d\kappa>0$), and it is unstable otherwise. This means that $\mathcal{B}_a$ is stable, while $\mathcal{B}_b$ is unstable.

We can also determine the STSs stability by analyzing their Hamiltonian function
\begin{equation}
	H=E\left[-\frac{a^2}{2}\rho+\frac{1}{2a^2}\left(1-\frac{E\nu\eta}{6\pi}\right)+\frac{7}{15}d_4\eta^4\right].  
\end{equation}
Once evaluated at the equilibrium STS solutions of Eqs.~(\ref{eq1}) and (\ref{eq2}), the Hamiltonian becomes just a function of $E$, and we may write $H_e\equiv H(E)$. This function is plotted in Fig.~\ref{fig2}(b).
According to the Lyapunov stability criteria \cite{parrarivas2023dynamics}, all STS solutions on the $\mathcal{B}_a$ branch minimize $H_e$: therefore, they are stable. However, those on $\mathcal{B}_b$ are unstable, since they maximize $H_e$. Thus, both stability criteria lead to the same result. 

The question that remains to be answered is if such approximate solutions, and their predicted stability, describe accurately enough the STS solutions of Eq.~(\ref{GPE}). To bring light to this, we performed full 3D numerical simulations of Eq.~(\ref{GPE}) by using advanced numerical algorithms based on a split-step predictor-corrector scheme \cite{frolkovic_numerical_1990}. 
To solve this initial value problem, we consider as initial condition the approximate analytical STS solution (\ref{ansatz1}), together with Eqs.~(\ref{eq1}) and (\ref{eq2}). The outcome of these computations is illustrated in Fig.~\ref{fig3}. Figure~\ref{fig3}(a) compares the analytically predicted peak intensity of the STS (see green line) with the numerically obtained associated values (see blue dots). In the latter, the circles and error bars represent the time-averaged intensity values, and the corresponding standard deviation for stable states. Stable STS are center steady states of Eq.~(\ref{GPE}) \cite{parrarivas2023dynamics}: therefore, they are neutrally stable. This means that any small perturbation leads to breathing oscillation around such points. Therefore, in practice, a steadily propagating STS is difficult to achieve. This fact may explain why, while the agreement between the variational approach and numerics is quite good for low values of energy, it worsens when increasing $E$. The $z$-propagation of a STS is illustrated in Fig.~\ref{fig3}(c).2, together with the $z$-evolution of its peak intensity [see Fig.~\ref{fig3}(c).1]. For large values of $E$, we find that the STSs undergo wave collapse (see red shadowed area) before the fold $f$, as it was the case in the pure quadratic regime \cite{parrarivas2023dynamics}. To compare the latter scenario with the former one, we plot approximate and numerically obtained $I_p$ values for pure quadratic dispersion in  Fig.~\ref{fig3}(b). This comparison shows that pure quartic dispersion significantly delays the appearance of wave collapse, by increasing by more than twice the $E$-range of STS existence. 

To summarize, in this letter we have reported on the formation of pure quartic STSs in GRIN waveguides. We show that pure quartic dispersion influences positively the propagation of STS, by leading to a significant widening of their energy stability range, and to the partial arrest of spatiotemporal collapse.

	
	This work was supported by European Research Council (740355), Marie Sklodowska-Curie Actions (101023717,101064614), Sapienza University of Rome Additional Activity for MSCA (EFFILOCKER), Ministero dell’Istruzione, dell’Università e della Ricerca (R18SPB8227).
	
	
	

\bibliographystyle{ieeetr}
\bibliography{Refs}
\end{document}